\documentclass[preprint]{aastex}
\usepackage{psfig}

\newcommand{\et}{et~al.\ }
\newcommand{\Ha}{H$\alpha$}

\newcommand\ltwid{\mathrel{\raise.5ex\hbox{$<$\kern-.75em\lower1.2ex\hbox{$\sim$}}}}

\shorttitle{COMPOSITE SEYFERT 2 X-RAY SPECTRUM}
\shortauthors{MORAN ET AL.}

\begin{document}

\title{A Composite Seyfert 2 X-ray Spectrum: Implications for the
       Origin of the Cosmic X-ray Background}

\author{Edward C.\ Moran\altaffilmark{1,2}, Laura E.\ Kay\altaffilmark{3},
        Marc Davis\altaffilmark{2}, Alexei V.\ Filippenko\altaffilmark{2},
        and Aaron J.\ Barth\altaffilmark{4}}

\altaffiltext{1}{Chandra Fellow.}

\altaffiltext{2}{Department of Astronomy, University of California,
                 Berkeley, CA 94720-3411.}

\altaffiltext{3}{Department of Physics and Astronomy, Barnard College,
                 Columbia University, New York, NY 10027.}

\altaffiltext{4}{Harvard-Smithsonian Center for Astrophysics, 60 Garden
                 Street, Cambridge, MA 02138.}

\begin{abstract}
We present a composite 1--10 keV Seyfert~2 X-ray spectrum, derived from
{\sl ASCA\/} observations of a distance-limited sample of nearby galaxies.
All 29 observed objects were detected.  Above $\sim$~3 keV, the composite
spectrum is inverted, confirming that Seyfert~2 galaxies as a class have
the spectral properties necessary to explain the flat shape of the cosmic
X-ray background spectrum.  Integrating the composite spectrum over redshift,
we find that the total emission from Seyfert~2 galaxies, combined with
the expected contribution from unabsorbed type~1 objects, provides an
excellent match to the spectrum and intensity of the hard X-ray background.
The principal uncertainty in this procedure is the cosmic evolution of
the Seyfert~2 X-ray luminosity function.  Separate composite spectra for
objects in our sample with and without polarized broad optical emission
lines are also presented.
\end{abstract}

\newpage
\section{Introduction}

Extensive observational effort over the past two decades has provided
compelling evidence that various types of radio-quiet active galactic
nuclei (AGNs), despite the diversity of their apparent properties, are
essentially similar objects that {\it look\/} different because of the
presence or absence of dense circumnuclear material along our line of sight
(Antonucci 1993).  The primary effect of such material is to obscure the
innermost regions of an active nucleus; this gives rise to a type~2 optical
spectrum which consists of relatively narrow permitted and forbidden emission
lines and lacks the very broad Balmer lines that characterize the spectra
of unobscured (type~1) AGNs.  In addition, the dense material preferentially
attenuates the soft X-ray emission produced by an AGN, flattening its
X-ray spectrum at lower energies (Awaki \et 1991).  This unified picture of
AGNs has thus allowed significant progress to be made toward a resolution
of the most enduring problem in X-ray astronomy---the origin of the cosmic
X-ray background (XRB)---by providing a means by which the integrated
emission from intrinsically steep-spectrum AGNs might be able to match the
flat observed spectrum of the background.  Models based on AGN unification
have shown that absorbed active nuclei may indeed account for both the
spectrum and intensity of the XRB over~a broad range of energies
(Setti \& Woltjer 1989; %Madau \et 1994; Comastri \et 1995).
Madau, Ghisellini, \& Fabian 1994; Comastri et al.~1995).

These models, however, rely on several crucial assumptions: (1) that
{\it all\/} type~2 AGNs are obscured type~1 objects; (2) that the absorption
column densities of active galaxies have a specific distribution; and (3)
that the X-ray luminosity function of absorbed AGNs evolves with redshift
in a particular way.  None of these conditions has been fully verified
observationally.  To explore the issue of whether or not all type~2 AGNs
are hidden type~1 objects, we have conducted an optical spectropolarimetric
survey of a well-defined sample of 31 nearby Seyfert 2 galaxies (Moran
\et 2000).  Broad \Ha\ emission has been detected in 13 of the objects,
suggesting that at least 42\% of Seyfert~2 galaxies are indeed obscured
Seyfert~1s.\footnote{Re-observation of part of our sample has revealed
faint polarized broad emission lines in NGC~5347 and NGC~5929, raising the
number of hidden Seyfert~1s in our sample from 11 to 13.  Details will
be presented in a future paper.}  Recent progress has also been made
toward a determination of the column density distribution in Seyfert
galaxies (e.g., Bassani \et 1999; Risaliti, Maiolino, \& Salvati 1999).
An alternative approach for assessing the contribution of absorbed AGNs to
the XRB would be to measure their integrated broadband spectral properties
within a known volume.  A composite spectrum based on the summed emission
of such objects would simultaneously represent several characteristics of
the Seyfert~2 population that are free parameters in AGN models of the XRB.
In this {\it Letter}, we present for the first time a composite 1--10 keV
Seyfert~2 X-ray spectrum, derived from a large, unbiased sample of nearby
objects, and discuss its implications for the origin of the XRB.

\section{X-ray Observations and Analysis}

Our study is based on the distance-limited sample of Ulvestad \& Wilson
(1989, hereafter UW89; see also Moran \et 2000), which contains 31 type~2
Seyfert galaxies with recession velocities $cz < 4600$ km~s$^{-1}$ and
declinations $\delta > -45^{\circ}$.  As of mid-1998, 22 objects in the
sample had been (or were scheduled to be) observed with the {\sl ASCA\/}
and/or {\sl BeppoSAX\/} satellites.  We were awarded time for the nine
remaining targets in {\sl ASCA\/} Cycle~7 .  Unfortunately, two of the
observations scheduled prior to this were given low-priority status and
were never carried out.  The galaxies lacking X-ray data (NGC 5283 and
NGC 5728) have average radio and optical luminosities compared to the rest
of the sample, and their absence here is not expected to have a significant
impact on our results.  Although a number of objects were observed with
both {\sl ASCA\/} and {\sl BeppoSAX}, all 29 were observed with {\sl ASCA},
so we have limited our study to that dataset.  In addition, because of
the consistency of the response and field of view of {\sl ASCA}'s Gas
Imaging Spectrometers (GIS2 and GIS3), we have utilized data obtained with
those instruments only.

We have employed ``rev2''-processed and screened {\sl ASCA\/} data.\footnote
{See {\tt http://heasarc.gsfc.nasa.gov/docs/asca/ascarev2.html} for details
regarding rev2 data processing.}  Exposures for the majority of the sample
range between $\sim$~35 ks and 45 ks.  All of the galaxies were observed at
either the ``1-CCD'' or ``2-CCD'' positions in the field of view, so all
objects are
approximately the same distance off-axis in the GIS images.  Two galaxies,
NGC 1068 and Mrk 3, were observed twice each, but in both cases the additional
data were obtained at different off-axis locations and have therefore been
omitted.  Source counts were extracted within a $6'$ radius region centered
on the galaxy for most objects; $4'$--$5'$ radius regions were used for the
weakest sources.  The background was estimated within a $13'$ radius circle
centered on the optical axis, excluding an $8'$ region around the source
and small areas around resolved background sources, if present.  Table~1
lists the galaxies observed, along with their distances (assuming $H_0 =
75$ km s$^{-1}$ Mpc$^{-1}$) and GIS exposure times.  Also listed are the
signal-to-noise ratios achieved with the combined GIS2 and GIS3 data in the
1--10 keV band.  As the table indicates, all objects were detected above a
significance of $\sim 4\sigma$.  Note, however, that over a third of the
sample is detected at or below $10\sigma$; thus, it would be impossible
to determine an accurate distribution of absorption column densities with
these data.  Source and background spectra with 128 channels were created
for each object from the extracted events, and effective area files were
generated for each of the source spectra.

Our objective is to construct a composite spectrum that is representative
of the local Seyfert~2 population, i.e., one in which the contribution of
individual objects is governed by their apparent luminosities, not their
observed fluxes.  Thus, we have multiplied the exposure times in the FITS
headers of the background-subtracted spectra of each galaxy by the factor
$(60/d)^2$, where $d$ is the distance of the source in Mpc from Table~1.
The modified GIS count rates are listed in the final column of Table~1.
Summing the modified count rates of the objects (using the FTOOLS task
{\tt mathpha}), we obtain a composite spectrum that represents what we
would observe if the objects were all located at 60~Mpc.  The sums were
carried out for the GIS2 and GIS3 data separately, yielding nearly
identical results.  The GIS2 and GIS3 sums were then averaged to produce
a final composite 1--10 keV Seyfert~2 X-ray spectrum.  Poisson errors were
propagated at each step in the procedure.  The individual effective area
files were averaged with the {\tt addarf} task.  We note that if the
objects {\it were\/} all 60~Mpc away, the data obtained would correspond
to a total exposure time of $8.5 \times 10^6$~s, which illustrates why it
would be difficult to repeat this experiment for a more distant sample
of comparable size.

\section{The Composite Seyfert~2 X-ray Spectrum}

The composite 1--10 keV Seyfert 2 spectrum, derived from the combined
emission of 29 galaxies, is displayed in Figure~1.  There is a considerable
amount of steep-spectrum emission below 3 keV, which was not anticipated
in the original Seyfert~2 XRB models.  Above 3~keV, however, the composite
spectrum is inverted, with strong Fe~K$\alpha$ line emission at 6.4 keV
(rest).  Again, the objects in our sample are included on the basis of
their distances and optical classifications, not their X-ray properties.
The composite spectrum thus verifies that Seyfert~2 galaxies {\it as a
class\/} have the X-ray spectral properties necessary to explain the flat
shape of the XRB spectrum---something which cannot be claimed for any other
extragalactic population.  Examination of the modified count rates listed
in Table~1 indicates that a single source, NGC~4507, contributes $\sim 30$\%
of the flux of the composite spectrum and that eight other objects (Mrk 3,
NGC 262, 424, 788, 1068, 3081, 3281, and 5135) are responsible for an
additional $\sim 50$\%.  Among these, NGC 424, 1068, and 5135 contribute
mainly in the soft X-ray band.  The others, which are the strongest sources
above 4~keV, all have ``Compton-thin'' spectra with absorption column
densities (inferred with the model described below) of $N_{\rm H}$ =
2--8 $\times 10^{23}$ cm$^{-2}$.

We have modeled the composite spectrum as the sum of a Gaussian line and
two power laws, one which is heavily absorbed and one which is not.  An
acceptable fit is obtained if we require the energy index $\alpha$ of the
power-law components to have the same value.  Such a model corresponds to
the ``partial scattering'' scenario, in which a single continuum component
is observed both in transmission through dense material and via scattering
along a less obscured line of sight.  However, for a handful of Seyfert~2
galaxies that have been studied with {\sl BeppoSAX}, this model underpredicts
the observed flux in the 20--30 keV range (e.g., Cappi \et 1999), probably
because it fails to account for the presence of Compton-reflected emission.
Limited to the 1--10 keV band, we are unable to constrain the strength of a
reflection component in the composite spectrum.  A double power-law
model in which the energy indices are independent also provides an acceptable
fit to the composite spectrum; the fit is shown in Figure~1.  This model may
not provide the correct physical description of the emission, but when it is
applied to the {\sl ASCA\/} spectra of individual Seyfert~2s observed with
{\sl BeppoSAX\/} (e.g., several of the objects studied by Cappi \et 2000)
it {\it does\/} accurately predict their measured fluxes in the 20--30 keV
range.  This is important because our estimate of the contribution of
Seyfert~2 galaxies to the intensity of the XRB (\S~4) necessarily involves
an extrapolation of the composite spectrum to higher energies.

We have compiled composite spectra separately for objects in our sample
which do and do not exhibit evidence for polarized broad optical emission
lines (see Table~1).  These are displayed in Figure~2.  Although there are
differences in detail between these spectra, they have similar qualitative
features: a declining continuum in the soft X-ray band which rises sharply
at harder energies, and a prominent Fe~K$\alpha$ line.  Thus, objects
lacking optical evidence for a hidden broad-line region---at least the
X-ray--brightest of them---display the characteristics of obscured type~1
Seyferts in the 1--10 keV range.  Note that the steep soft X-ray component
is relatively stronger in Seyfert~2s that do {\it not\/} have polarized
optical emission lines, which casts some doubt on the possibility that
the soft X-ray flux in these objects is mainly scattered emission.

\section{Seyfert 2 Galaxies and the X-ray Background}

As we have discussed previously (Moran \et 2000), the UW89 objects comprise
a reasonably complete, distance-limited sample of classical Seyfert~2 galaxies.
Since they occupy a known volume, the composite spectrum we have derived is
effectively an unbiased measure of the local Seyfert~2 volume emissivity
(luminosity per unit volume) as a function of X-ray energy.  This can be
integrated over redshift using the following equation to provide an estimate
of the sky brightness at energy $E$ due to such objects:
$$I_{\rm E} = {c \over {4 \pi H_0}}
\int\displaylimits_{0}^{z_{\rm max}} {{\rho(E')\; e(z)\; dz} \over 
{(1 + z)\, [\Omega_{\rm M}\, (1 + z)^3 + \Omega_{\Lambda}]^{1/2} }}\; ,$$
\noindent
where $\rho$ is the volume emissivity at $E' = E(1 + z)$, $e(z)$ is a term
that describes the evolution of the volume emissivity with redshift, and
$\Omega_{\rm M}$ and $\Omega_{\Lambda}$ are the matter and vacuum-energy
density parameters, respectively.
Since we have ``moved'' all of the galaxies to $d = 60$ Mpc,
the composite spectrum shown in Figure~1 (which is the sum of their fluxes)
is converted to a volume emissivity by multiplying by $4 \pi d^2$ and dividing
by 0.6 times the volume of a 60 Mpc sphere.\footnote{The sample does not
cover the Galactic plane ($|b| < 20^{\circ}$) or the south celestial cap
($\delta < -45^{\circ}$), hence the factor of 0.6.}  Although the evolution
term $e(z)$ is unmeasured for type~2 AGNs, it has been characterized for
type~1 objects in the soft X-ray band by Miyaji, Hasinger, \& Schmidt
(2000).  For simplicity, we have adopted their pure-density evolution
model, which has the following form: $e(z) = (1 + z)^{4.4}$ for $z \le 1.6$,
and $e(z) = e(1.6)\, [(1 + z) / 2.6]$ for $z > 1.6$ (T.~Miyaji 2001, private
communication).  This evolution model can account for all of the soft XRB
with type~1 AGNs, so our results should be considered an upper limit for
the contribution of type~2 Seyferts to the hard XRB.  Finally, we have
adopted an $\Omega_{\rm M} = 0.3$, $\Omega_{\Lambda} = 0.7$ cosmology for
our calculations (Garnavich \et 1998; Balbi \et 2000; Hanany \et 2000)
and an upper integration limit of $z_{\rm max} = 6$ (AGNs have been detected
out to this redshift).

The results of the integration in the 1--10 keV band are displayed in Figure~3.
As anticipated (see Gilli \et 1999), the integration has smoothed out all of
the bumps and wiggles in the composite Seyfert~2 spectrum.  Assuming that
steep-spectrum ($\alpha = -0.8$) type~1 AGNs produce $\sim$~60\% of the XRB
at 1~keV---consistent with the findings of the {\sl ROSAT\/} deep survey
of the ``Lockman Hole'' (Schmidt \et 1998)---the combined contribution of
Seyfert~1 and Seyfert~2 galaxies agrees exceptionally well with the observed
spectrum and intensity of the XRB (from Gendreau \et 1995) above $\sim$~3 keV.
Our calculations are not particularly sensitive to the adopted cosmology.
For example, if we assume $\Omega_{\rm M} = 1.0$ and $\Omega_{\Lambda} =
0.0$, the Seyfert~2 contribution we obtain has the same overall shape as
that shown in Figure~3 and about two-thirds of the integrated intensity.

In conclusion, the spectrum of the summed broadband X-ray emission of
nearby Seyfert~2 galaxies and its integration over redshift provide
strong support for the hypothesis that absorbed AGNs play a significant
(if not dominant) role in the production of the hard X-ray background.
Some important details need to be investigated further before we can
consider the XRB problem to be solved, namely, the average spectral
properties of Seyfert~2 galaxies above 10~keV and the cosmic evolution
of the Seyfert~2 X-ray luminosity function.  In addition, if optically
classified Seyfert~2 galaxies produce most of the hard XRB, as Figure~3
suggests, then the majority of the flat-spectrum sources being detected
in deep {\it Chandra\/} images should be Seyfert~2s.  Currently, this
does not appear to be the case---a significant fraction of faint, hard
{\it Chandra\/} sources have starlight-dominated optical spectra with
emission lines that are weak or absent (Barger \et 2001; Hornschemeier
\et 2001).  It is possible that the Seyfert~2 X-ray luminosity function
evolves less rapidly than we have assumed, which would leave room for
other types of AGNs---objects so obscured there is no optical evidence
of their activity---to emerge at faint flux levels.  Alternatively, the
limitations of ground-based observing have almost certainly affected our
ability to classify faint {\it Chandra\/} sources.  The majority of
these sources are quite distant, so in spectroscopic observations most of
the light from their host galaxies is collected along with the nuclear
emission.  At the low spectral resolution that is typically employed,
this starlight dilution can have a profound impact on the appearance of
the nuclear emission-line spectrum.  A powerful demonstration of this
effect has been presented by Kennicutt (1992).  Thus, the faint, hard
{\it Chandra\/} source population may {\it not\/} be significantly
different from the nearby Seyfert~2 galaxies we have investigated here.

\acknowledgments

We are grateful to T.~Miyaji for helpful discussions about the AGN
X-ray luminosity function, and to the {\sl ASCA\/} Cycle~7 peer review
panel for approving a substantial amount of satellite time for this
project. We also thank our referee for a number of insightful comments.
E.C.M.\ is supported by NASA through Chandra Fellowship grant PF8-10004
awarded by the Chandra X-ray Center, which is operated by the Smithsonian
Astrophysical Observatory for NASA under contract NAS 8-39073.  L.E.K.\
acknowledges the support of the NSF through CAREER grant AST-9501835.
A.J.B.\ is supported by a postdoctoral fellowship from the
Harvard-Smithsonian Center for Astrophysics.  We also acknowledge NASA
grant NAG 5-3556 and a Guggenheim Fellowship to A.V.F.

\begin{center}
\begin{deluxetable}{lcccc}
\tablewidth{0pt}
\tablecaption{The Seyfert 2 Sample}
\tablehead{\colhead{Galaxy} &
           \colhead{$d$\tablenotemark{a}} &
           \colhead{Exp.\tablenotemark{b}} &
           \colhead{{~SNR}\tablenotemark{c}} &
           \colhead{{CR}\tablenotemark{d}}}
\startdata
IC  3639\tablenotemark{e}&  43.7  & 36.2  & ~10.2 &  $2.4 \times 10^{-3}$\\
MCG $-$05-18-002~~~~~    &  23.1  & 34.6  & ~12.9 &  $1.3 \times 10^{-3}$\\
MCG $+$01-27-020         &  46.8  & 41.6  & ~~8.0 &  $2.2 \times 10^{-3}$\\
Mrk 3\tablenotemark{e}   &  54.0  & 36.7  & ~44.3 &  $3.2 \times 10^{-2}$\\
Mrk 1066                 &  48.1  & 36.2  & ~12.6 &  $4.3 \times 10^{-3}$\\
NGC 262\tablenotemark{e} &  60.1  & 45.5  & ~41.4 &  $3.2 \times 10^{-2}$\\
NGC 424\tablenotemark{e} &  46.6  & 34.2  & ~20.6 &  $8.7 \times 10^{-3}$\\
NGC 591\tablenotemark{e} &  60.7  & 40.9  & ~~7.3 &  $2.8 \times 10^{-3}$\\
NGC 788\tablenotemark{e} &  54.4  & 40.2  & ~35.5 &  $2.3 \times 10^{-2}$\\
NGC 1068\tablenotemark{e}&  14.4  & 96.6  & 173.6 &  $1.1 \times 10^{-2}$\\
NGC 1358                 &  53.8  & 36.2  & ~~9.9 &  $3.6 \times 10^{-3}$\\
NGC 1386                 &  16.9  & 37.6  & ~18.2 &  $9.9 \times 10^{-4}$\\
NGC 1667                 &  60.7  & 42.3  & ~~4.2 &  $1.5 \times 10^{-3}$\\
NGC 1685                 &  60.4  & 34.0  & ~~3.8 &  $1.6 \times 10^{-3}$\\
NGC 2273\tablenotemark{e}&  28.4  & 36.9  & ~20.0 &  $2.3 \times 10^{-3}$\\
NGC 3081\tablenotemark{e}&  32.5  & 36.7  & ~39.8 &  $8.8 \times 10^{-3}$\\
NGC 3281                 &  42.7  & 18.4  & ~19.1 &  $1.1 \times 10^{-2}$\\
NGC 3982                 &  27.2  & 40.6  & ~10.3 &  $9.1 \times 10^{-4}$\\
NGC 4117                 &  17.0  & 38.0  & ~25.3 &  $1.2 \times 10^{-3}$\\
NGC 4388\tablenotemark{e}&  16.8  & 28.1  & ~37.1 &  $3.8 \times 10^{-3}$\\
NGC 4507\tablenotemark{e}&  47.2  & 40.4  & 102.9 &  $9.0 \times 10^{-2}$\\
NGC 4941                 &  ~6.4  & 36.0  & ~13.9 &  $1.0 \times 10^{-4}$\\
NGC 5135                 &  54.9  & 44.0  & ~18.6 &  $9.2 \times 10^{-3}$\\
NGC 5347\tablenotemark{e}&  36.7  & 36.7  & ~~6.6 &  $9.8 \times 10^{-4}$\\
NGC 5643                 &  16.9  & 41.5  & ~29.6 &  $1.4 \times 10^{-3}$\\
NGC 5695                 &  56.4  & 24.6  & ~~4.6 &  $1.9 \times 10^{-3}$\\
NGC 5929\tablenotemark{e}&  38.5  & 42.2  & ~22.1 &  $5.8 \times 10^{-3}$\\
NGC 6890                 &  31.8  & 34.7  & ~~7.0 &  $5.7 \times 10^{-4}$\\
NGC 7672                 &  53.5  & 39.3  & ~~4.7 &  $1.3 \times 10^{-3}$
\enddata
\tablenotetext{a}{Distances (in Mpc) from Tully (1988).  For the
                  more distant sources ($cz > 3000$ km~s$^{-1}$),
                  we assume $d = cz/H_0$.}
\tablenotetext{b}{Average exposure time (in ks) per detector.}
\tablenotetext{c}{Signal-to-noise ratio for the combined 1--10 keV GIS data.}
\tablenotetext{d}{Modified 1--10 keV GIS count rate (in ct s$^{-1}$), i.e.,
                  that which would be observed if the objects were at 60 Mpc.}
\tablenotetext{e}{Has polarized broad optical emission lines.}
\end{deluxetable}
\end{center}

\begin{figure}
\begin{center}
\centerline{\psfig{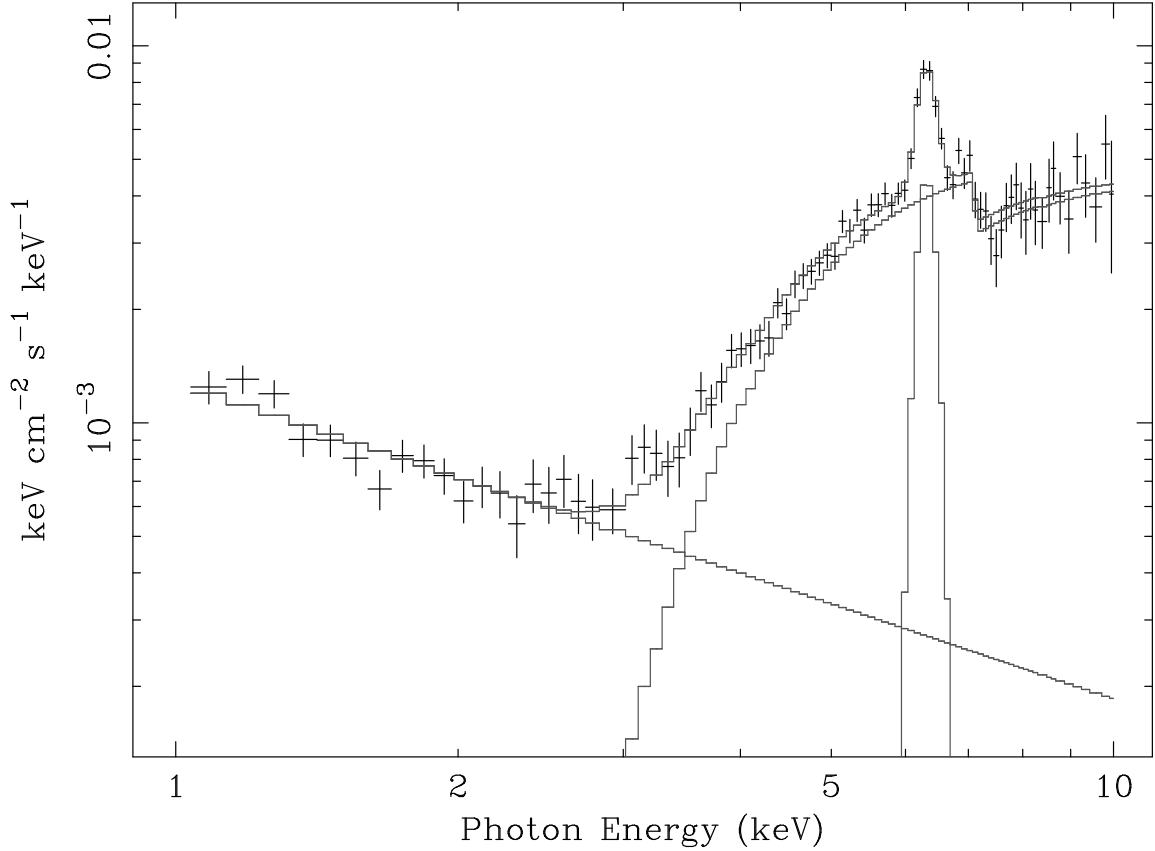}}
\vskip 0.2truein
\caption{The composite Seyfert~2 X-ray spectrum, fitted with a double
power-law model.  An unabsorbed component (energy index $\alpha = -0.84$)
dominates below 3 keV; at higher energies, the spectrum is dominated by
a heavily absorbed component ($N_{\rm H} = 3.0 \times 10^{23}$ cm$^{-2}$,
$\alpha = -0.46$).  The equivalent width of the Fe K$\alpha$ line at
6.4 keV is 420 eV.  In the 1--10 keV band, the spectrum of the XRB is modeled
as an $\alpha = -0.4$ power law.}
\end{center}
\end{figure}

\begin{figure}
\begin{center}
\centerline{\psfig{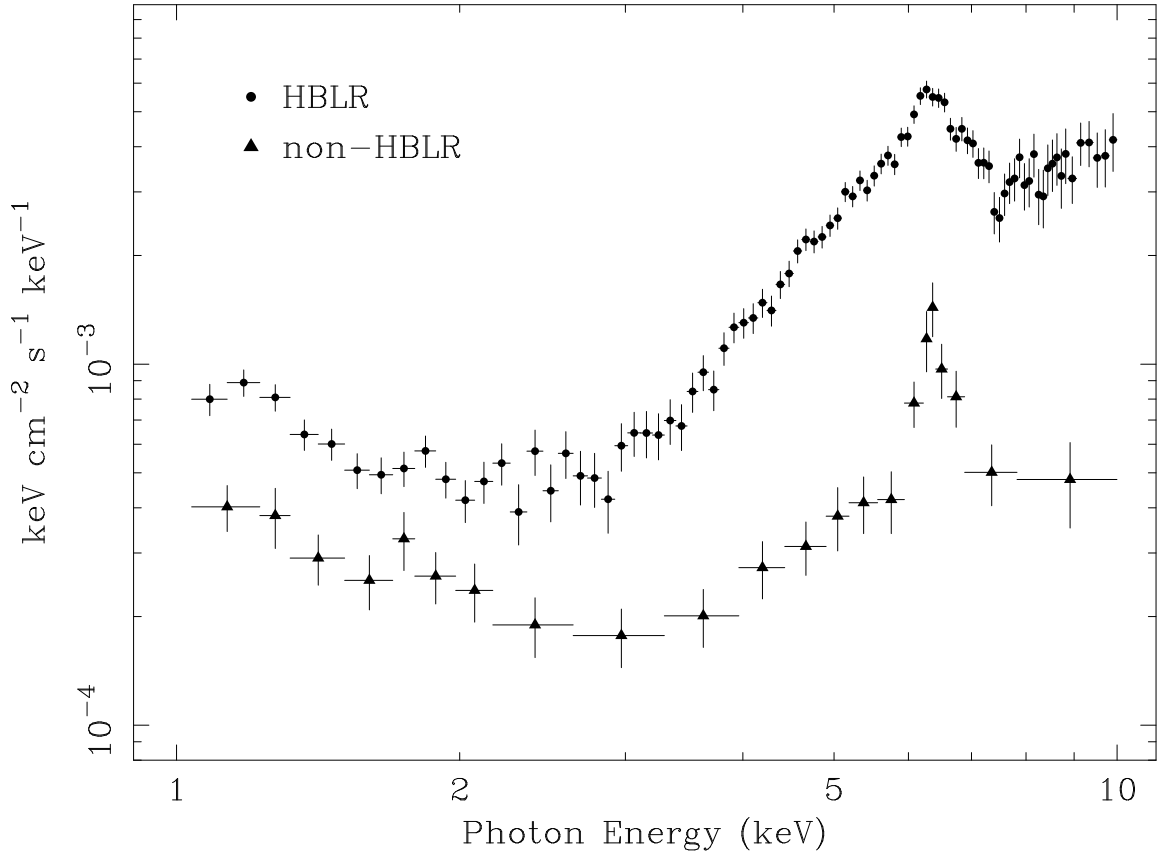}}
\vskip 0.2truein
\caption{Composite X-ray spectra for the 13 Seyfert~2 galaxies in our
sample that have polarized broad optical emission lines, which indicate
the presence of a hidden broad-line region (HBLR), and for the 16 objects
that do not (non-HBLR).}
\end{center}
\end{figure}

\begin{figure}
\begin{center}
\centerline{\psfig{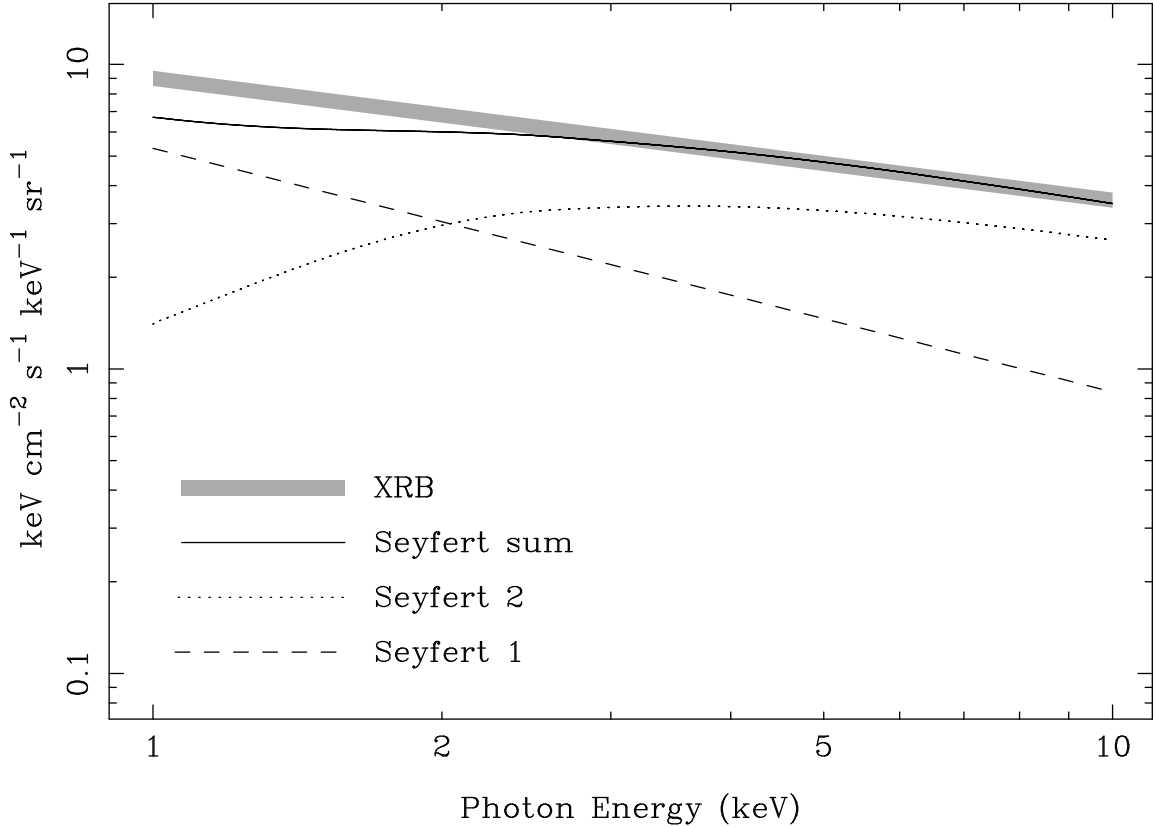}}
\vskip 0.2truein
\caption{X-ray sky brightness due to Seyfert~2 galaxies ({\it dotted
line}), obtained via integration of the composite spectrum over redshift.
Note that much of the structure in the composite spectrum has been
smoothed out.  The contribution expected from steep-spectrum type~1
AGNs ($\sim$~60\% of the XRB at 1 keV; Schmidt \et 1998) is also shown
({\it dashed line}).  The combined emission from type~1 and type~2
Seyfert galaxies ({\it solid line}) compares well with the spectrum of
the XRB measured by Gendreau \et (1995), especially at $E > 3$ keV.
The discrepancy between the XRB spectrum and the Seyfert sum below 2 keV
could represent the contributions of groups and clusters of galaxies
and star-forming galaxies, which are not included in this simple model.}
\end{center}
\end{figure}

\end{document}